\documentclass[aps,nofootinbib,superscriptaddress,twocolumn,prd,10pt,floatfix]{revtex4-1}

\usepackage{amsmath}
\usepackage{mathrsfs}
\usepackage{graphicx}
\usepackage{amssymb}
\usepackage{comment}

\usepackage{hyperref}
\usepackage[dvipsnames]{xcolor}

\hypersetup{
    colorlinks=true,
    linkcolor=red,
    citecolor=blue,
}

\newcommand{\figref}[1]{Fig.~\ref{#1}}

\newcommand{\wn}{w_{\rm f}}
\newcommand{\ADE}{{\rm ADE}}

\begin{document}

\title{Acoustic Dark Energy: Potential Conversion of the Hubble Tension}
\author{Meng-Xiang Lin}
\affiliation{Kavli Institute for Cosmological Physics, Department of Astronomy \& Astrophysics, Enrico Fermi Institute, The University of Chicago, Chicago, IL 60637, USA}
\author{Giampaolo Benevento}
\affiliation{Dipartimento di Fisica e Astronomia ``G. Galilei'', Universit\`a degli Studi di Padova, via Marzolo 8, I-35131, Padova, Italy}
\affiliation{INFN, Sezione di Padova, via Marzolo 8, I-35131, Padova, Italy}
\affiliation{Kavli Institute for Cosmological Physics, Department of Astronomy \& Astrophysics, Enrico Fermi Institute, The University of Chicago, Chicago, IL 60637, USA}
\author{Wayne Hu}
\affiliation{Kavli Institute for Cosmological Physics, Department of Astronomy \& Astrophysics, Enrico Fermi Institute, The University of Chicago, Chicago, IL 60637, USA}
\author{Marco Raveri}
\affiliation{Kavli Institute for Cosmological Physics, Department of Astronomy \& Astrophysics, Enrico Fermi Institute, The University of Chicago, Chicago, IL 60637, USA}

\begin{abstract}
We discuss the ability of a dark fluid becoming relevant around the time of matter radiation equality to significantly relieve the tension between local measurements of the Hubble constant and CMB inference, within the $\Lambda$CDM model.
We show that the gravitational impact of acoustic oscillations in the dark fluid balance the effects on the CMB and result in an improved fit to CMB measurements themselves while simultaneously raising the Hubble constant.
The required balance favors a model where the fluid is a scalar field that converts its potential to kinetic energy around matter radiation equality which then quickly redshifts away.
We derive the requirements on the potential for this conversion mechanism and find that a simple canonical scalar with two free parameters for its local slope and amplitude robustly improves the fit to the combined data by $\Delta\chi^2 \approx 12.7$ over $\Lambda$CDM.
We uncover the CMB polarization signatures that can definitively test this scenario with future data.
\end{abstract}

\maketitle

\section{Introduction} \label{Sec:introduction}

The $\Lambda$CDM model of cosmology has been tested by an extensive number of independent probes, showing its general robustness and remarkable ability to explain a wide range of
observational data with only six parameters.
Despite its indubitable success, $\Lambda$CDM seems to fail in reconciling distance-redshift measurements when anchored at high redshift by cosmic microwave background (CMB) anisotropies to the same measurements anchored at low redshift by the local distance ladder.

This discrepancy is commonly quantified as tension between  the inferences for the Hubble constant ($H_0$), and its statistical significance has been steadily increasing with increasing experimental precision.
The most recent local estimate of the Hubble constant places its value at $H_0=74.03\pm1.42$  km\,s$^{-1}$\,Mpc$^{-1}$~\cite{Riess:2019cxk}, showing a $4.4 \sigma$ tension with the value inferred by the Planck 2018 CMB data, $H_0= 67.4 \pm 0.5$  km\,s$^{-1}$\,Mpc$^{-1}$, assuming the $\Lambda$CDM model~\cite{Aghanim:2018eyx}. \\

This tension is mainly a discrepancy between the anchors for the absolute distance scale rather 
than an indicator of missing physics between the anchors.   
Once anchored at one end, the same ladder of intermediate redshift measurements from baryon acoustic oscillations (BAOs) to supernovae Type IA (SN) predict the anchor at the other end, leaving little room for missing cosmological physics in between (for a recent assessment and discussion, see Refs.~\cite{Bernal:2016gxb,Aylor:2018drw} and references therein).   
On the high redshift side, the anchor is the CMB sound horizon $r_s$.
Under $\Lambda$CDM, the shapes of the CMB acoustic peaks calibrate the sound speed and all of the energy densities of species relevant around recombination and thus determine the physical scale of $r_s$.  Measurements of its angular  scale in the CMB then fix the remaining parameter, the cosmological constant or equivalently $H_0$.   
Even beyond $\Lambda$CDM, this measurement determines the distance to recombination and anchors the inverse distance ladder from which $H_0$ can again be inferred so that even the most general dark energy or modified gravity model can only moderately alter its value~\cite{Zhao:2017cud,Lin:2018nxe,Raveri:2019mxg,Benevento:2019}.

Altering the high redshift anchor requires modifying cosmological physics at recombination.   Adding extra energy density  raises the expansion rate before recombination and lowers the sound horizon $r_s$.   
For example, an extra sterile neutrino or other dark radiation has long been considered as a possible solution~\cite{Wyman:2013lza,Dvorkin:2014lea,Leistedt:2014sia,Ade:2015rim,Lesgourgues:2015wza,Adhikari:2016bei,DiValentino:2016hlg,Canac:2016smv,Feng:2017nss,Oldengott:2017fhy,Lancaster:2017ksf,Kreisch:2019yzn}.   
However  such a component would also change the driving of the  acoustic oscillations and damping scale~\cite{Hu:1996vq}, which is now disfavored by increasingly precise CMB data, leaving little ability to raise $H_0$ 
(see Refs.~\cite{Raveri:2017jto,Aghanim:2018eyx} for recent assessments).   

The problem with the damping scale arises because these additions affect the background expansion like radiation.
As pointed out in Ref.~\cite{Poulin:2018cxd}, this problem can be avoided by making the dark component only important transiently  near the epoch of recombination.
Specifically, Ref.~\cite{Poulin:2018cxd} introduces a component of so-called ``early dark energy" (EDE) where a scalar field oscillates anharmonically around the minimum of a periodic potential and finds that the Hubble tension can be efficiently relieved.   
On the other hand, Ref.~\cite{Agrawal:2019lmo} finds that for a monomial EDE model, which coincides with the periodic potential~\cite{Poulin:2018cxd} at the minimum, the Hubble tension is only partially relieved.   
Due to the behavior of its perturbations, these EDE scenarios change the amplitudes and phases of the CMB acoustic peaks in complex ways, leading to questions as to the robustness of this method for relieving the Hubble tension.

To build a more robust method to relieve the Hubble tension, we study  the general phenomenology of perturbations in a dark fluid which similarly becomes transiently important, here around matter-radiation equality.   
Since its impact on the CMB comes through the gravitational effects of its own acoustic oscillations, we call this species acoustic dark energy (ADE) and in particular uncover the critical role its sound speed plays in relieving the Hubble tension.   
We find that the sound speed must vary with the equation of state in the background in a manner consistent with the conversion of potential to kinetic energy for a minimally coupled scalar with a general kinetic term~\cite{ArmendarizPicon:2000ah}.
Unlike the oscillatory EDE models, once released from Hubble drag, the scalar remains kinetic energy dominated until it redshifts away.
Indeed for a simple canonical kinetic term, this allows for $H_0 = 70.60 \pm 0.85$ with a better fit than $\Lambda$CDM even for the CMB alone and a better total $\chi^2$ by 12.7 for 2 extra parameters.  
This method is also robust and can be exactly realized in a wide class of potentials.    
We provide both the required conditions on the potential and explicit examples.

This paper is organized as follows.
In \S\ref{Sec:datasets}, we introduce the ADE fluid model, its parameters, and the data sets that we use in the analysis.
In \S\ref{Sec:ADEphenom}, we discuss the phenomenological impacts of ADE, especially its sound speed, on acoustic driving and CMB polarization. 
In \S\ref{Sec:conversion}, we show that ADE models that can relieve the Hubble tension correspond to scalars that convert potential to kinetic energy suddenly upon Hubble drag release
and construct a canonical scalar model as proof of principle.
In \S\ref{Sec:previous}, we discuss the relation to the previous work and we conclude in \S\ref{Sec:discussion}.

\section{Methodology} \label{Sec:datasets}

Acoustic dark energy is defined to be a perfect dark fluid and is specified by its background equation of state $w_{\ADE}=p_\ADE/\rho_\ADE$ and  rest frame sound speed $c_s^2$~\cite{Hu:1998kj}.  The latter is only equivalent to the  adiabatic sound speed $\dot p_\ADE/\dot \rho_\ADE$ for a barotropic fluid so that in the general case the acoustic phenomenology of linear \ADE\ sound waves,
which we shall see is crucial for relieving the Hubble tension, is defined independently of the background.

In order to have a transiently important ADE contribution, we model the ADE equation of state as
\begin{equation} \label{eqn:eos}
1+w_{\ADE}(a) = \frac{1+\wn }{[1+(a_c/a)^{3(1+\wn )/p}]^{p}} \,.
\end{equation}
The ADE component therefore changes its equation of state around  a scale factor $a=a_c$ from $w_{\ADE}=-1$ to $\wn$. 
Additionally, $p$ controls the rapidity of this transition, with small values corresponding to sharper transitions.
Since this parameter does not qualitatively change our results, we use $p=1/2$ unless otherwise specified.   
We shall see in \S \ref{Sec:conversion} that this corresponds to a simple quadratic potential for scalar field ADE.
This is a generalization of the background of the EDE model~\cite{Poulin:2018cxd}, discussed in
\S \ref{Sec:previous}, where $p=1$ and the fluid description is approximate.
  
The ADE background energy density is fully specified once its normalization is fixed, since $w_\ADE$ determines its evolution.
Defining the ADE fractional energy density contribution
\begin{equation}
f_\ADE(a) = \frac{\rho_\ADE(a)}{\rho_{\rm tot}(a)} \,,
\end{equation}
we choose $f_c=f_\ADE(a_c)$ as the normalization parameter.

The behavior of ADE perturbations is determined by their rest frame sound speed $c_s^2(a,k)$ which is, for an effective fluid, a function of both time and scale~\cite{Hu:1998kj}.  In the context of a perfect fluid with a linear dispersion relation, it is scale independent.   In particular this holds for K-essence  scalar field models~\cite{ArmendarizPicon:2000ah}, when treated exactly instead of in a time-averaged approximation.   We shall return to this point in \S \ref{Sec:previous}.

The equations of motion for ADE acoustic oscillations depend only on the value of $c_s^2$, not its time derivative.
Since the impact of ADE on cosmological observables is  extremely localized in time due to the parametrization of $w_{\ADE}$, we fix $c_s^2$ to be a constant, effectively its value at $a_c$.
In \S \ref{Sec:conversion}, we construct K-essence models where $c_s^2$ is strictly constant as a proof of principle, but our analysis holds more generally if we interpret the constant $c_s^2$ as matching a suitably averaged evolving one.  

In our most general case, the ADE model is therefore characterized by four parameters $\{ \wn,  a_c, f_c, c_s^2 \}$ once $p$ is fixed.
When varying these parameters we impose flat, range bound priors:
$-4.5\leq \log_{10}a_c \leq -3.0$, $0\leq f_c \leq 0.2$, $0\leq \wn \leq 3.6$ and $0\leq c_s^2 \leq 1.5$.
We shall later see that the Hubble tension can be relieved by varying just two of these four parameters, fixing $c_s^2=\wn=1$, corresponding to models where the ADE is a canonical scalar that converts its energy density from potential to kinetic around matter radiation equality (see \S \ref{sec:canonmodel}). We refer to this particular ADE model as cADE.

The full cosmological model also includes the six $\Lambda$CDM parameters: the cold dark matter density is characterized by $\Omega_c h^2$, baryon density by $\Omega_b h^2$, the angular size of the sound horizon by $\theta_{s}$, the optical depth to reionization by $\tau$, and the initial curvature spectrum by its normalization at $k=0.05$ (wavenumbers throughout are in units of Mpc$^{-1}$, which we drop when no confusion should arise), $A_s$ and tilt $n_s$.
These have the usual non-informative priors. We fix the sum of neutrino masses to the minimal value (e.g., Ref.~\cite{Long:2017dru}).
We modify the CAMB~\cite{Lewis:1999bs} and CosmoMC~\cite{Lewis:2002ah} codes to include all the models that we discuss, following Ref.~\cite{Hu:1998kj}. 
We sample the posterior parameter distribution until the Gelman-Rubin convergence statistic~\cite{gelman1992} satisfies $R-1<0.02$ or better unless otherwise stated.

For the principal cosmological data sets, we use the publicly available Planck 2015 measurements of the CMB temperature and polarization power spectra at large and small angular scales and the CMB lensing potential power spectrum in the multipole range $40 \leq \ell \leq 400$~\citep{Ade:2015xua,Aghanim:2015xee,Ade:2015zua}.
To expose the Hubble tension, we combine this with the latest measurement of the Hubble constant, $H_0=74.03\pm1.42$ (in units of km\,s$^{-1}$\,Mpc$^{-1}$ here and throughout)~\citep{Riess:2019cxk}.
To these data sets we add the Pantheon Supernovae sample~\cite{Scolnic:2017caz} and BAO measurements from BOSS DR12~\cite{Alam:2016hwk}, SDSS Main Galaxy Sample~\cite{Ross:2014qpa}, and 6dFGS~\cite{Beutler:2011hx}.
These datasets prevent resolving the Hubble tension by modifying the dark sector only between recombination and the very low redshift universe~\cite{Aylor:2018drw}.  

Our baseline configuration thus contains: CMB temperature, polarization and lensing, BAO, SN and $H_0$ measurements.  
Unless otherwise specified all of our results will be for this combined data set.
We include all the recommended parameters and priors describing systematic effects for these data sets. 

As we shall see, the CMB polarization data provide an important limitation on the ability to raise $H_0$ and future polarization data can provide a definitive test of the ADE models that alleviate the Hubble tension.   We therefore also consider the joint data set without CMB polarization data.  We refer to this data set as -POL.

\begin{table*}[!ht]
\setlength{\tabcolsep}{12pt}
\centering
\begin{tabular}{@{}cccccc@{}}
\toprule
Model                  & $\Lambda$CDM                           & cADE       & ADE             \\
\toprule                                                                                                               
$100\theta_{\rm MC}$   & 1.04115 (1.04110$\pm$0.00028)   & 1.04062 (1.04064$\pm$0.00031)  & 1.04072 (1.04081$\pm$0.00035)  \\
$\Omega_bh^2$          & 0.02246 (0.02241$\pm$0.00014)   & 0.02267 (0.02271$\pm$0.00022)  & 0.02270 (0.02263$\pm$0.00022)  \\
$\Omega_ch^2$          & 0.1170 (0.1174$\pm$0.0009)      & 0.1268 (0.1268$\pm$0.0032)     & 0.1274 (0.1242$\pm$0.0032)     \\
$\tau$                 & 0.082 (0.075$\pm$0.012)         & 0.064 (0.064$\pm$0.012)        & 0.064 (0.067$\pm$0.013)        \\
$\ln(10^{10}A_s)$      & 3.092 (3.079$\pm$0.022)         & 3.078 (3.078$\pm$0.023)        & 3.080 (3.081$\pm$0.023)        \\
$n_s$                  & 0.9726 (0.9701$\pm$0.0039)      & 0.9833 (0.9833$\pm$0.0065)     & 0.9873 (0.9832$\pm$0.0071)     \\
\colrule
$f_c$                  & -                              & 0.082 (0.082$\pm$0.025)        & 0.086 (0.079$\pm$0.033)        \\
$\log_{10}a_c$         & -                              & -3.45 (-3.46$\pm$0.06)         & -3.52 (-3.50$\pm$0.15)         \\
$\wn$                  & -                              & 1 ({\rm fixed})               & 0.87 (1.89$\pm$0.86)           \\
$c_{s}^2$              & -                              & 1 ({\rm fixed})               & 0.86 (1.07$\pm$0.25)           \\
\colrule                                                                                                                
$H_0$                  & 68.58 (68.35$\pm$0.42)          & 70.57 (70.60$\pm$0.85)         & 70.81 (70.20$\pm$0.88)         \\
$\Delta\chi^2_{\rm tot}$ & 0                            & -12.7                         & -14.1                       \\
\botrule
\end{tabular}
\caption{ \label{tab:parameters}
Maximum likelihood (ML) parameters and  constraints for the $\Lambda$CDM model, the canonical Acoustic Dark Energy (cADE) model, and the general ADE model. $\Delta\chi^2_{\rm tot} = -2\Delta \ln {\cal L}$ reflects the ratio between the maximum likelihood value and that of $\Lambda$CDM for the joint data. 
}
\end{table*}

\section{ADE Phenomenology} \label{Sec:ADEphenom}

In this section we discuss the phenomenology and observational implications of ADE 
and their dependence on its parameters.

At the background level, the addition of ADE increases the total energy density before recombination that changes the expansion history lowering the sound horizon $r_s$.  
This changes the calibration of distance measures not only for the CMB but also the whole inverse distance ladder through BAO to SN.   
Given the precise angular measurements of the sound horizon $\theta_s$, the inverse distance ladder scale is reduced and hence the inferred $H_0$ rises.   

The prototypical example of this method for relieving the Hubble tension is an extra sterile neutrino that is at least mildly relativistic at recombination.
Neutrinos, however, do not provide a good global solution~(e.g., Ref.~\cite{Raveri:2017jto}) since they behave as free-streaming radiation before recombination and therefore change the phase of the CMB acoustic oscillations as well as the CMB damping scale, the distance a photon random walks through the ionized plasma before recombination, approximately as $\lambda_D \propto r_s^{1/2}$~\cite{Hu:1996vq}.
A more general dark fluid, on the other hand,  can reduce the fraction of the dark component vs.~normal radiation before matter radiation equality, allowing the two scales to change in a proportional way~\cite{Poulin:2018cxd}.  

Beyond these background effects, ADE and other dark sector candidates for relieving the
Hubble tension,  gravitationally drive photon-baryon acoustic oscillations changing the amplitudes and phases of the CMB peaks (e.g., Ref.~\cite{Raveri:2017jto,Lin:2018nxe}).
ADE perturbations undergo their own acoustic oscillations under its sound horizon, leading to novel CMB driving phenomenology.
As detailed in Table~\ref{tab:parameters}, this modified phenomenology leads to a maximum likelihood (ML)  solution 
with $\wn \approx c_s^2$ and $H_0=70.81$ which improves $\Delta  \chi^2_{\rm tot}=-14.1$ over ML $\Lambda$CDM, and more generally  a finite ADE fraction $f_c$ is detected at $2.4\sigma$.
In the next section, we focus on the details of these physical effects.  We then
address the impact of Planck polarization data and the ability of future polarization data to test the ADE solutions to the Hubble tension.

Finally, note that although we do not consider measurements of the amplitude of local structure here, the ML and  constraints for ADE are $\sigma_8\Omega_m^{1/2}$=0.4623  (0.4573$\pm$0.0073), whereas for $\Lambda$CDM with CMB TT only has nearly the same ML but larger errors 0.466 (0.466$\pm$0.013).     
For our combined data set, $\Lambda$CDM gives the value 0.4488 (0.4486$\pm$0.0056), where the ML is lower since raising $H_0$ in $\Lambda$CDM lowers $\sigma_8\Omega_m^{1/2}$ unlike in ADE.  
If the tension with weak lensing measurements of the amplitude increases in $\Lambda$CDM in the future, it will disfavor not only $\Lambda$CDM but these ADE models as well.

\subsection{Acoustic Driving} \label{Sec:AcousticDriving}

Under the sound horizon or Jeans scale of the ADE, its density perturbations acoustically oscillate rather than grow, leading to changes in the decay of the Weyl potential $(\Psi+\Phi)/2$.   
This decay drives CMB acoustic oscillations, and the ADE impact is especially important for modes that enter the CMB sound horizon near $a_c$, roughly $k=0.04$ in the ML ADE model from Table~\ref{tab:parameters}.  
The excess decay is countered by raising the cold dark matter through $\Omega_c h^2$ since it remains gravitationally unstable on the relevant scales.

\begin{figure}[!ht]
\centering
\includegraphics[width=\columnwidth]{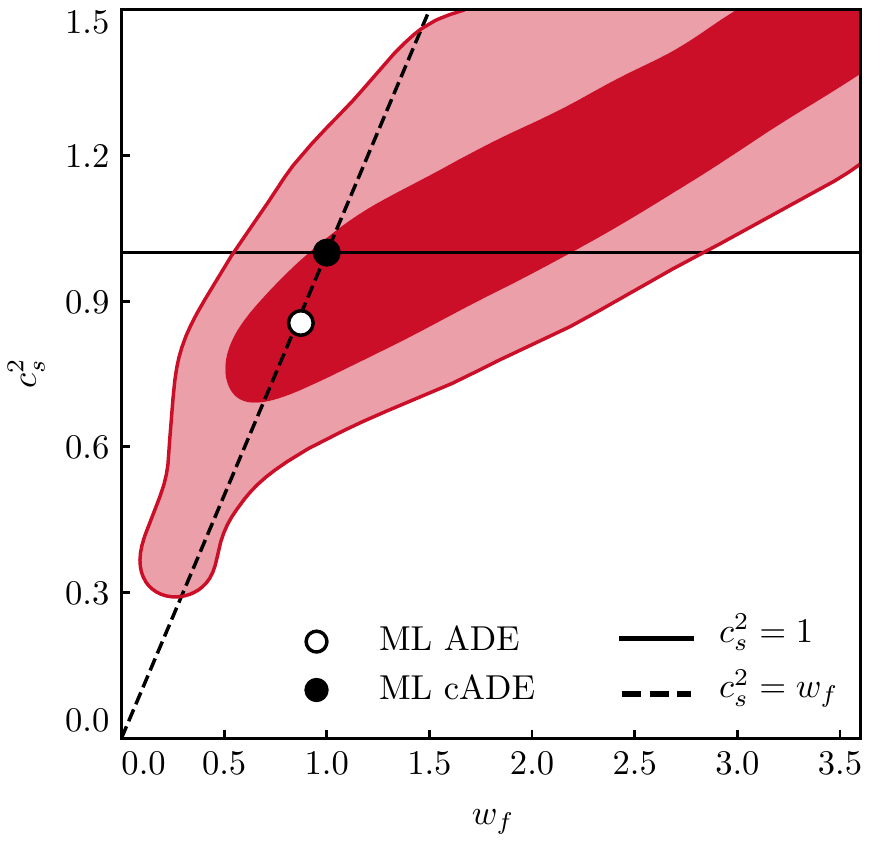}
\caption{
\label{fig:wn-cs2} 
The joint marginalized distribution of the ADE parameters $c_s^2$ and $\wn$, obtained using our combined datasets. 
The darker and lighter shades correspond respectively to the 68\% C.L. and the 95\% C.L.
The markers indicate the maximum likelihood values for ADE (solid circle) from Tab.~\ref{tab:parameters} and the intersection between canonical models $c_s^2=1$ (solid line) and models which convert potential to kinetic energy at the transition $c_s^2=\wn$ (dashed line),~i.e. $c_s^2 = \wn =1$ (open circle) as in cADE.
}
\end{figure}

For ADE, at the parameter level, this effect is controlled by the sound speed $c_s$ in conjunction with the equation of state $w_\ADE(a)$ through $\wn$.
These two parameters are hence strongly correlated, as shown in Fig.~\ref{fig:wn-cs2}, reflecting degenerate effects on the CMB when they are raised or lowered together.  Near the ML
solution, this requires $\wn \approx c_s^2$.

\begin{figure}[!ht]
\centering
\includegraphics[width=\columnwidth]{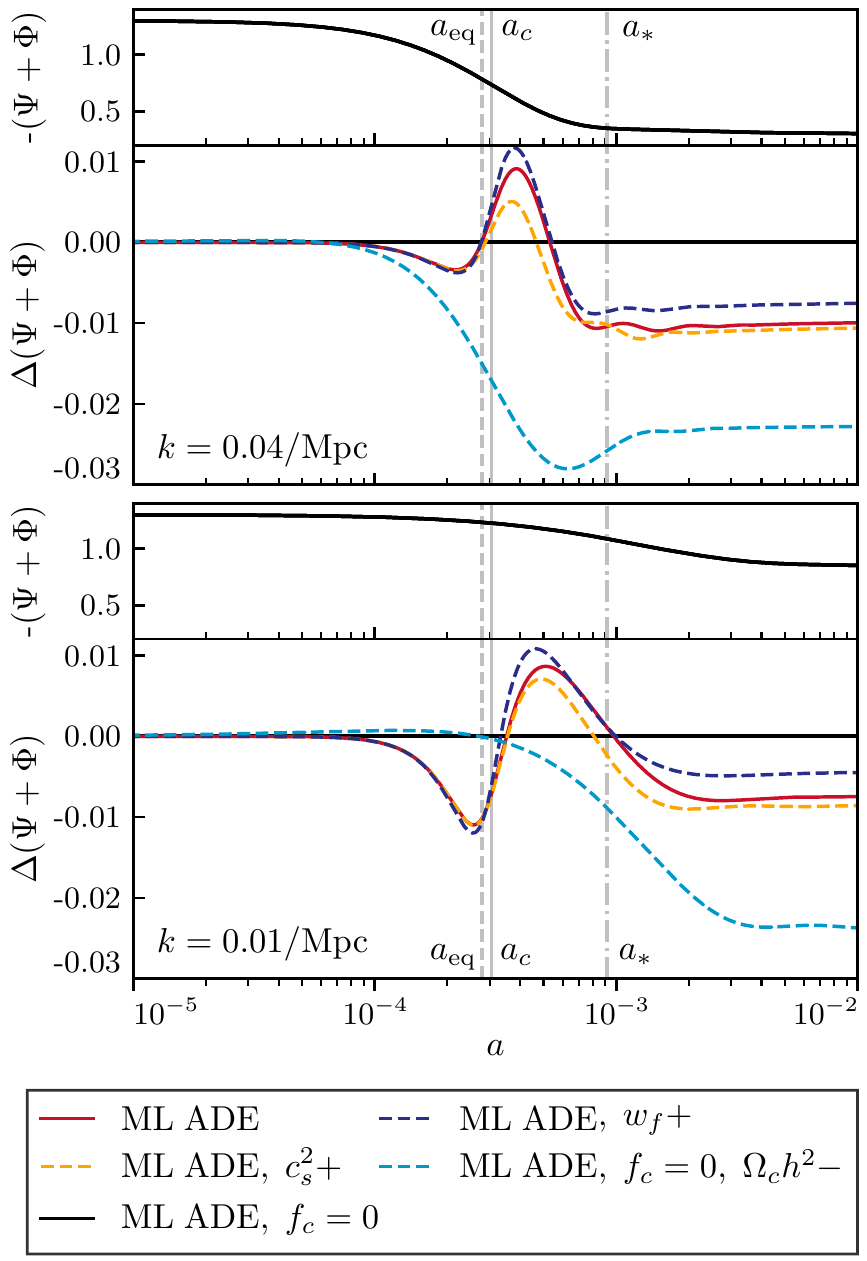}
\caption{ \label{fig:Weyl}
The Weyl potential evolution of the ML ADE model from Table~\ref{tab:parameters} for two modes: $k=0.01$ and $0.04\;{\rm Mpc^{-1}}$. 
Lower subpanels show differences with respect to the  baseline value of Weyl potential for the $\Lambda$CDM parameters of ML ADE but with no ADE ($f_c=0$), as displayed in the upper subpanels.
Shown are ML ADE (red solid) and  parameter variations around it:
$c_s^2+$ (orange dashed) and $w_f+$ (dark blue dashed) mean $+0.4$ variations, while $\Omega_c h^2-$ (cyan dashed) means lowering it  to the ML $\Lambda$CDM value in Table~\ref{tab:parameters}.   Relevant temporal scales (matter-radiation equality $a_{\rm eq}$, ADE transition $a_c$ and recombination $a_*$) are shown with
vertical lines.
}
\end{figure}

We explore this degeneracy in  Fig.~\ref{fig:Weyl} by showing the evolution of the Weyl potential for this mode in the ML model 
(red) from Table~\ref{tab:parameters} relative to the same model with no ADE (ML, $f_c=0$) as a baseline (black).
The Weyl potential  is relatively suppressed at $a<a_c$, enhanced at $a \sim a_c$ and suppressed again at $a \gg a_c$ due to ADE.   
The enhancement and subsequent suppression correspond to the  first acoustic compression extremum in the ADE density perturbation and the subsequent Jeans stabilization of the perturbations.  
The net impact is a reduction in the Weyl potential.   This reduction is compensated by raising $\Omega_c h^2$.    
For comparison we also show the difference in reverting the value of $\Omega_c h^2$ in the baseline model to the ML $\Lambda$CDM value (cyan dashed).   
Since $a_c \sim a_{\rm eq}$, ADE becomes important around the same epoch when radiation driving  has the maximal impact on the shape of the CMB acoustic peaks.
Along with other adjustments in  $\Lambda$CDM parameters, in particular $n_s$ and $\Omega_b h^2$, these effects compensate for each  other. 

\begin{figure}[!ht]
\centering
\includegraphics[width=\columnwidth]{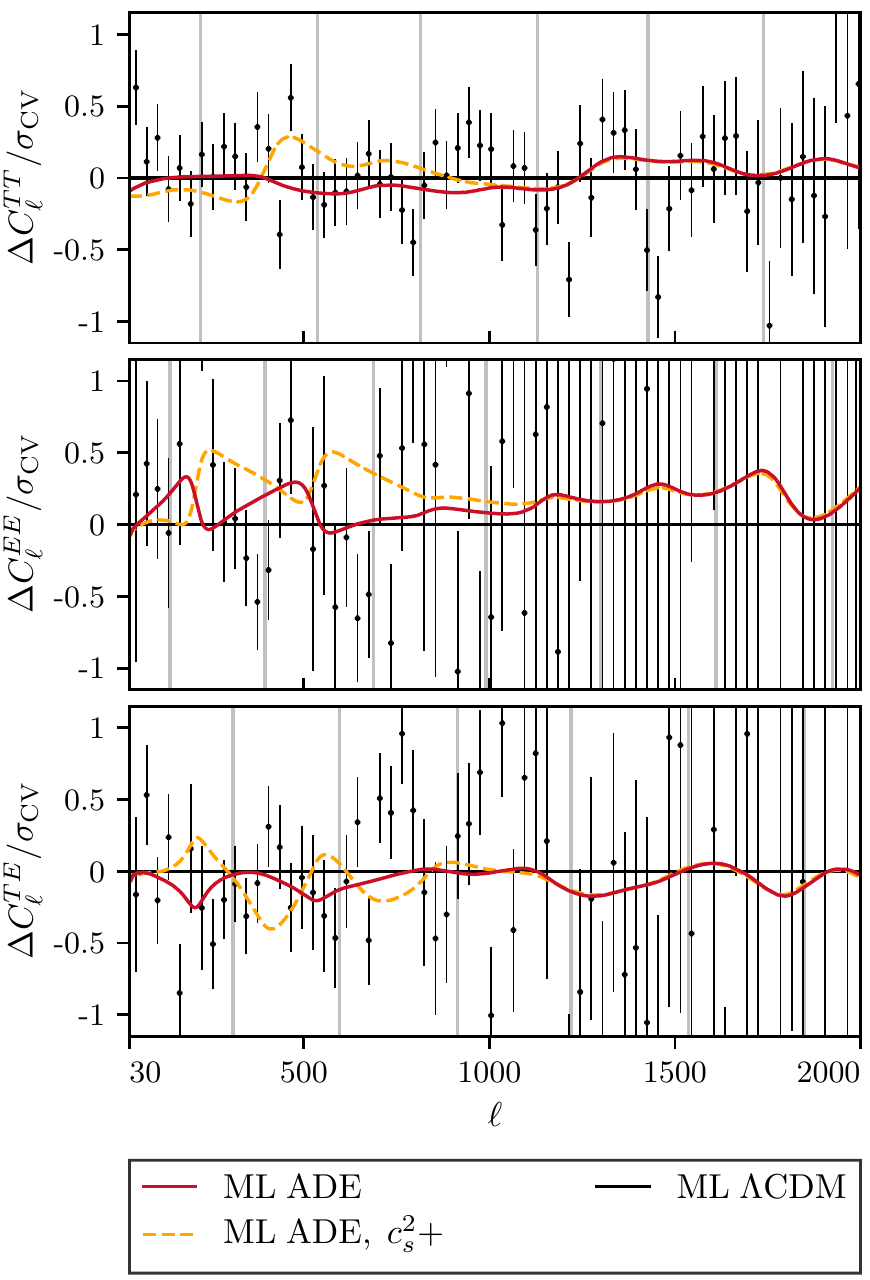}
\caption{
\label{fig:residual}
The CMB model and Planck data residuals with respect to the ML $\Lambda$CDM model.  
Shown are the ML ADE model (red solid) and a $\Delta c_s^2=+0.4$ variation on it (orange dashed, see Fig.~\ref{fig:Weyl}).
The upper, middle, and bottom panels correspond to TT, EE, and TE residuals respectively
and vertical lines denote their peaks in the  ML $\Lambda$CDM model.   
}
\end{figure}

This compensation  leaves the CMB acoustic peaks nearly unchanged despite raising $H_0$ from 68.58 to 70.81.  
In Fig.~\ref{fig:residual}, we show the data and model residuals relative to the $\Lambda$CDM ML in Table~\ref{tab:parameters}.
As we can see, this effect does not exacerbate the oscillatory residuals in the data, where the acoustic peaks (vertical lines) are suppressed relative to troughs,  which would occur if
 $H_0$ were raised in $\Lambda$CDM.   Note that the residuals are
scaled to the cosmic variance per $\ell$-mode for the ML $\Lambda$CDM model as:
\begin{equation}
\sigma_{\rm CV} =
\begin{cases}
\sqrt{\frac{2}{2\ell+1}} C_\ell^{TT}, & TT \,;\\
\sqrt{\frac{1}{2\ell+1}}  \sqrt{ C_\ell^{TT} C_\ell^{EE} + (C_\ell^{TE})^2}, & TE \,;\\
\sqrt{\frac{2}{2\ell+1}}  C_\ell^{EE}, & EE \,. \\
\end{cases}
\end{equation}

We can better understand the origin of the ADE effects and their impact on the CMB by varying $\wn$ and $c_s^2$ independently.   Fig~\ref{fig:Weyl} (upper) also shows a $+0.4$ variation in each with other parameters
held fixed.
Increasing $c_s^2$ makes the ADE acoustic oscillations and
 Jeans stability occur earlier and as a consequence also cuts into the enhancement. Raising $\wn$ has two effects.   Before the first compression peak and above the CMB sound horizon, the comoving ADE density perturbation grows adiabatically so that its amplitude
 grows relative to the photons  approximately as $\delta_\ADE \propto (1+w_\ADE) \delta_\gamma$.   For $\wn>1/3$ this first causes a dip at $a \lesssim a_c$ and then an enhancement 
 in the Weyl potential for $a\gtrsim a_c$, especially approaching the first compression.   A larger $\wn$ then suppresses $f(a>a_c)$ which also causes a relative enhancement at $a>a_c$.
 Combined, these effects imply that for a fixed amount of driving of acoustic oscillations through the decay of the Weyl potential, raising $c_s^2$ should be compensated by 
 raising $\wn$.   This is the leading order degeneracy that we see in Fig.~\ref{fig:wn-cs2}.
 
In terms of the residuals, shown in Fig.~\ref{fig:residual}, a positive variation in $c_s^2$, when {\it not}
compensated by $\wn$, leads to a sharp TT feature around $\ell \sim 500$ near the second TT
 peak whereas along the degeneracy line the ML model TT residuals remain small.   The modes responsible
 for higher multipoles are not sensitive to ADE perturbation parameters since the Weyl decay that drives them occurred before
 the ADE became important $a\ll a_c$.
 
  The degeneracy is truncated at low $\wn$ in Fig.~\ref{fig:wn-cs2}. If $\wn <1/3$, the ADE redshifts slower than the radiation and thus has a large impact on the driving of
  CMB acoustic oscillations between $a_c$ and recombination which cannot be balanced by the same variations in $c_s^2$.

\subsection{CMB Polarization} \label{Sec:pol}
 
Even for the ML ADE model, the compensation between $\Omega_c h^2, c_s^2$, and $\wn$ is imperfect for  modes that enter the CMB sound horizon between $a_c$ and recombination.    These modifications leave distinct imprints on the polarization spectra that
already limit the ability of ADE to raise $H_0$ using the Planck data and, in the future, can 
definitively test this scenario.
   Polarization provides the cleanest signatures of driving  on these scales given that
it isolates the acoustic oscillations at recombination from the early integrated Sachs-Wolfe that smooths out its signatures in the TT spectrum. 

In Fig.~\ref{fig:Weyl} (lower), we show the Weyl potential evolution for such a mode, 
$k=0.01$  in the ML model.   While the qualitative behavior is similar to the higher $k$ mode, the balance between $\wn$, $c_s^2$ and $\Omega_c h^2$ changes.   
First, the impact of the change in $\Omega_c h^2$ is relatively higher since the ADE redshifts faster than radiation.  
Second, the impact of $\wn$ is also somewhat higher relative to $c_s^2$.   
These  changes lead to uncompensated sharp features in the polarization residuals.

For the Planck data, where measurements of the EE spectrum are still noisy, this makes the TE spectrum the most informative for these features (see Fig.~\ref{fig:residual}).   
In the $\Lambda$CDM model, this sensitivity provides an important constraint on $\Omega_c h^2$ and hence supporting evidence for a low $H_0$ from multipoles $\ell <1000$.   
The region around $\ell \sim 165$ (between the first EE and TT peaks)  is particularly important due to the $2\sigma$ low point compared to the best fit $\Lambda$CDM model shown in Fig.~\ref{fig:residual}~\cite{Obied:2017tpd}.  Note that in $\Lambda$CDM raising $H_0$ requires lowering $\Omega_c h^2$ which raises TE there, making the fit even worse.

In the \ADE\ ML model, the impact of raising $\Omega_c h^2$ lowers TE in this region providing a better fit to the data.
Even without direct $H_0$ data, the CMB data favor the ADE model (see Table~\ref{tab:chi2}).  However
raising $H_0$ further than the ADE ML would make TE too low at $\ell\lesssim 500$ for the
data. 
Indeed for the -POL dataset, the ML model allows for a larger Hubble
constant $H_0 = 72.27$ compared with $70.81$ for our joint dataset.

Raising $c_s^2$ has the impact of making the ADE more important for 
Weyl decay and counters the effect of raising $\Omega_c h^2$.   As we can see
from Fig.~\ref{fig:residual}, this has the effect of raising TE in this region and degrading the fit.
Thus the TE data are also important for disfavoring a canonical scalar field with $c_s^2=1$ if
$\wn$ is too low.  We shall see in \S \ref{Sec:previous} that this explains why previously considered models where a canonical
scalar field  oscillates in its potential must have its initial conditions set to avoid this region.

Finally, given the sharp features in the EE model residuals with up to $\sim 0.3-0.4$ amplitudes
relative to cosmic variance {\it per multipole}, all of these cases where $H_0$ is raised by 
dark components that also change the driving of the acoustic peaks can be tested to high
significance once EE measurements approach the precision of TT measurements today.

\section{Potential-Kinetic Conversion} \label{Sec:conversion}

The ADE phenomenology favored by the Hubble tension can be concretely and exactly 
realized in the
K-essence class of dark energy models, where the dark component is a perfect fluid represented
by a minimally coupled scalar field $\phi$ with a general kinetic term~\cite{ArmendarizPicon:2000ah}.  More specifically,
the class of constant sound speed $c_s$ models introduced in \S \ref{Sec:datasets} is given by the Lagrangian density
\cite{Gordon:2004ez}
\begin{equation}
P(X,\phi) =\left( \frac{X}{A}\right)^{\frac{1-c_s^2}{2 c_s^2}} X - V(\phi),
\label{eqn:Lagrangian}
\end{equation}
where the kinetic term involves  $X= -\nabla^\mu \phi \nabla_\mu \phi /2$ and $A$ is a constant density scale.   For a scalar with a canonical kinetic term $c_s^2=1$, and more generally $w_{\rm ADE} \rightarrow c_s^2$ if the kinetic term dominates,
whereas $w_{\ADE}\rightarrow -1$ if the potential $V(\phi)$ dominates.   
The fluid correspondence holds when $\nabla^{\mu}\phi$ remains always timelike;  then $c_s^2 = \delta p/\delta \rho$ in constant field
gauge or rest frame, where the momentum density of the field vanishes and the potential energy is spatially constant. 

The correlation shown in Fig.~\ref{fig:wn-cs2} implies that around the ML ADE model from Table~\ref{tab:parameters}, setting
$\wn = c_s^2$ provides a good fit to the combined data.  Since $w_\ADE\rightarrow -1$ for $a \ll a_c$ and $w_\ADE\rightarrow \wn$ for $a \gg a_c$, this suggests that the 
best fitting $P(X,\phi)$ models are those that suddenly convert
nearly all of their energy density from potential to kinetic at $a_c$.   
If we focus on a model that has such a potential to kinetic energy conversion feature, we have $c_s^2=\wn$ and the number of parameters of ADE reduces to three. The ML of this model gives $c_s^2=0.84$ , $H_0=70.84$ and $\Delta \chi^2_{\rm tot}=-14.1$ which is nearly identical to ML ADE with one fewer free parameter.

Indeed the preferred region includes $\wn=c_s^2=1$ which corresponds to the case of a canonical field, cADE.  
From Table~\ref{tab:parameters}, we see that the ML cADE model has only a marginally smaller 
$H_0=70.57$ for a $\Delta\chi^2_{\rm tot}=-12.7$ and more generally a higher significance
to the detection of a finite ADE fraction $f_c$ of $3.3\sigma$ given the smaller set of parameters.   
We shall now consider how to construct a corresponding potential $V(\phi)$.

\subsection{Canonical Conditions}

A canonical scalar which converts its potential to kinetic energy around $a_c$ provides
a simple, concrete example of ADE that alleviates the Hubble tension.   
To explicitly construct such a model that matches requirements on the two remaining 
quantities $a_c$ and $f_c$ we can determine the equivalent requirements for the
potential $V(\phi)$.

At $a\ll a_c$, the  $\phi$ field is stuck on its potential due to Hubble friction and rolls 
according to 
\begin{equation}
\frac{d\phi}{dN} \sim -\frac{V'}{H^2},
\label{eqn:Hubbledrag}
\end{equation}
where $N=\ln a$ denotes e-folds.
For the purposes of this qualitative discussion we drop factors of order unity.  After $a_c$, we want the field to be 
released from Hubble drag and  convert its potential energy to kinetic energy on the e-fold timescale $\Delta N\sim 1$. 
Defining $\phi_{c}= \phi(a_c)$ and linearizing the change
\begin{equation}
V(\phi) \approx V(\phi_{c} ) + V'(\phi_{c}) \Delta \phi.
\label{eqn:linearizedV}
\end{equation}
Therefore, around $\phi_{c}$, we want
\begin{equation}
\left(\frac{V'}{H}\right)^2 \gtrsim V,
\end{equation}
or in terms of  $f_{c} \sim V/\rho_{\rm tot}$,
\begin{equation}
 \epsilon_V f_c \gtrsim 1, \qquad \epsilon_V \equiv \frac{M_{\rm Pl}^2}{2}\left(\frac{V'}{V} \right)^2.
\end{equation}
This is the main condition for the potential to kinetic conversion. 

For the linearization in Eq.~(\ref{eqn:linearizedV}) to be valid in the sense of the second order term $\frac{1}{2}  (\Delta \phi)^2 V''$  not preventing the conversion, we also want
\begin{equation}
 V''(\phi_{c})  < -\frac{V'(\phi_{c}) }{\Delta\phi}
\end{equation}
so
\begin{equation} 
V'' \lesssim   H^2 .
\end{equation}
Putting these two criteria together, 
\begin{equation}
\eta_V  \lesssim  2 \epsilon_V, \qquad \eta_V \equiv M_{\rm pl}^2 \frac{V''}{V} \,,
\end{equation}
where we have restored a factor of 2 so as to match the well-known  condition for no tracking solution
to exist~\cite{Steinhardt:1999nw}.   Tracking potentials do not work since the scalar field follows an equation of state that is determined by the dominant component of the total energy density rather than
the kinetic energy dominated limit.  A similar derivation applies to the $c_s^2\ne 1$ case with a modification to the Hubble drag evolution (\ref{eqn:Hubbledrag})~\cite{Gordon:2004ez}.

Finally we want the field to maintain kinetic energy domination until its energy density has largely redshifted away.  This excludes models where the field oscillates around a minimum
and so is different from those in Refs. ~\cite{Poulin:2018cxd,Agrawal:2019lmo} as we shall discuss in the next section.  Furthermore, the fluid description is exact for our models whereas
it is only approximate for oscillatory models.

Thus our requirements on the potential are fairly generic and correspond to setting the amplitude and slope of the potential at the desired point of Hubble drag release, along
with the condition that the field remains kinetic energy dominated until most of the energy density has redshifted away.  A wide class of potentials can satisfy these requirements and we shall give concrete examples next.

\begin{figure}
\centering
\includegraphics[width=\columnwidth]{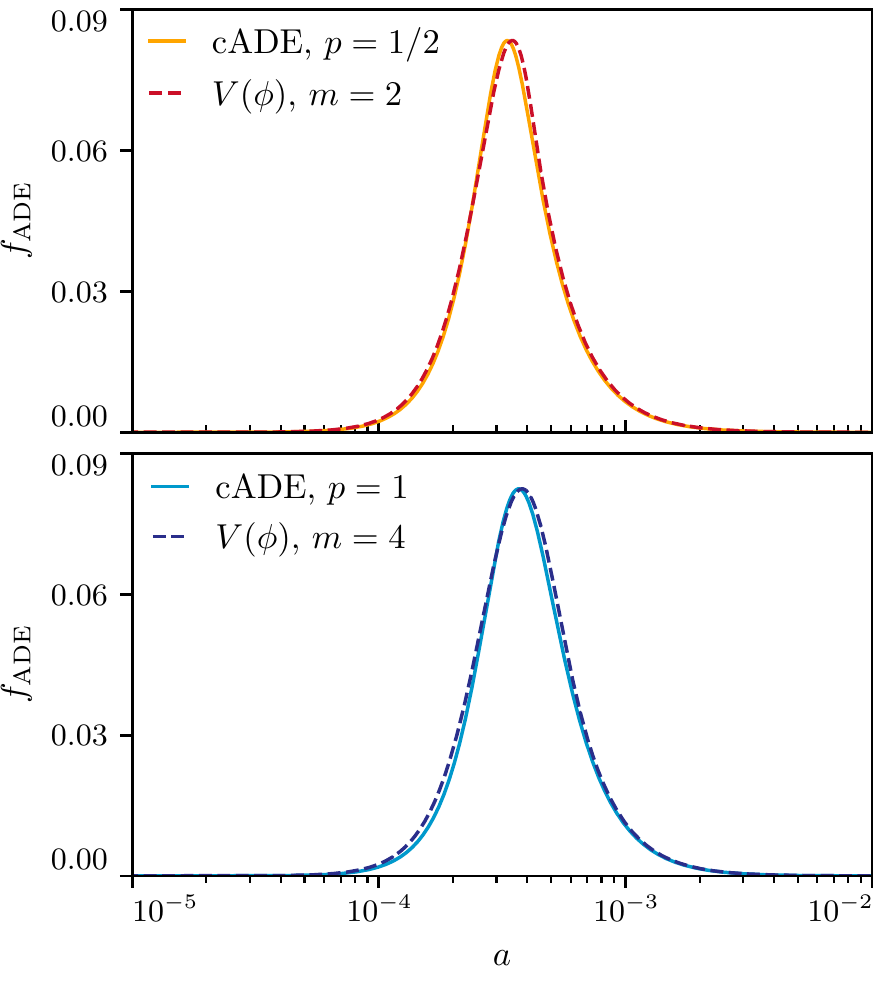}
\caption{ \label{fig:PotentialMatching}
Scalar field potential $V(\phi)$ match to the fractional ADE energy density $f_\ADE$ of the ML cADE parameters in Table~\ref{tab:parameters}.    Top: a locally quadratic potential with Eq.~(\ref{eqn:potential}) compared with $p=1/2$ in Eq.~(\ref{eqn:eos}); bottom: a locally
quartic potential vs $p=1$.  
}
\end{figure}

\subsection{Canonical Solution} \label{sec:canonmodel}
To make these considerations concrete, consider the class of potentials:
\begin{equation} \label{eqn:potential}
V(\phi) = 
\begin{cases} 
 A \, \phi^m,  &  \qquad \phi>0 \,, \\ 
 0,  & \qquad \phi \le 0 \,.
 \end{cases}
\end{equation} 
Then for $\phi >0$
\begin{equation}
2\epsilon_V=  \left( \frac{m}{\phi} \right)^2, \qquad \eta_V = \frac{m(m-1)}{\phi^2} \,,
\end{equation} 
and any $m>0$ satisfies $\eta_V<2 \epsilon_V$.   
The flat potential at $\phi \le 0$ prevents the kinetic energy from converting back to potential energy.
We choose $A$ and $\phi_{\rm initial}$ to give the desired $f_c$ and $a_c$. 

In Fig. \ref{fig:PotentialMatching} (upper), we show a worked example of this matching. 
We fix cosmological parameters to the ML ADE model in Table~\ref{tab:parameters} and  take a quadratic potential with $m=2$.  We find a good match to the
form of  Eq.~(\ref{eqn:eos}) with $p=1/2$. This motivates our fixed fiducial
choice  in \S \ref{Sec:datasets}.

To showcase the robustness of the potential to kinetic conversion mechanism for relieving the Hubble tension, we
also consider  a quartic potential $m=4$ (Fig.~\ref{fig:PotentialMatching}, lower).    The change in $f_\ADE(a)$ is itself small and, once a  shift in $a_c$ is absorbed, corresponds to a slight broadening of the transition.
Even this small change can be matched to the general ADE form of Eq.~(\ref{eqn:eos}) by adopting $p=1$.  
For the linear $m=1$ case,  $p\approx 0.1$ corresponding to a sharper transition.
We have also tested that various values of $p$ in this range provide comparable ML solutions
to our fiducial $p=1/2$ case.  Finally we have tested that the correspondence between $p$ and $m$ holds for non-canonical values of $c_s^2$ with the Lagrangian  (\ref{eqn:Lagrangian}).

\begin{figure}[!ht]
\centering
\includegraphics[width=\columnwidth]{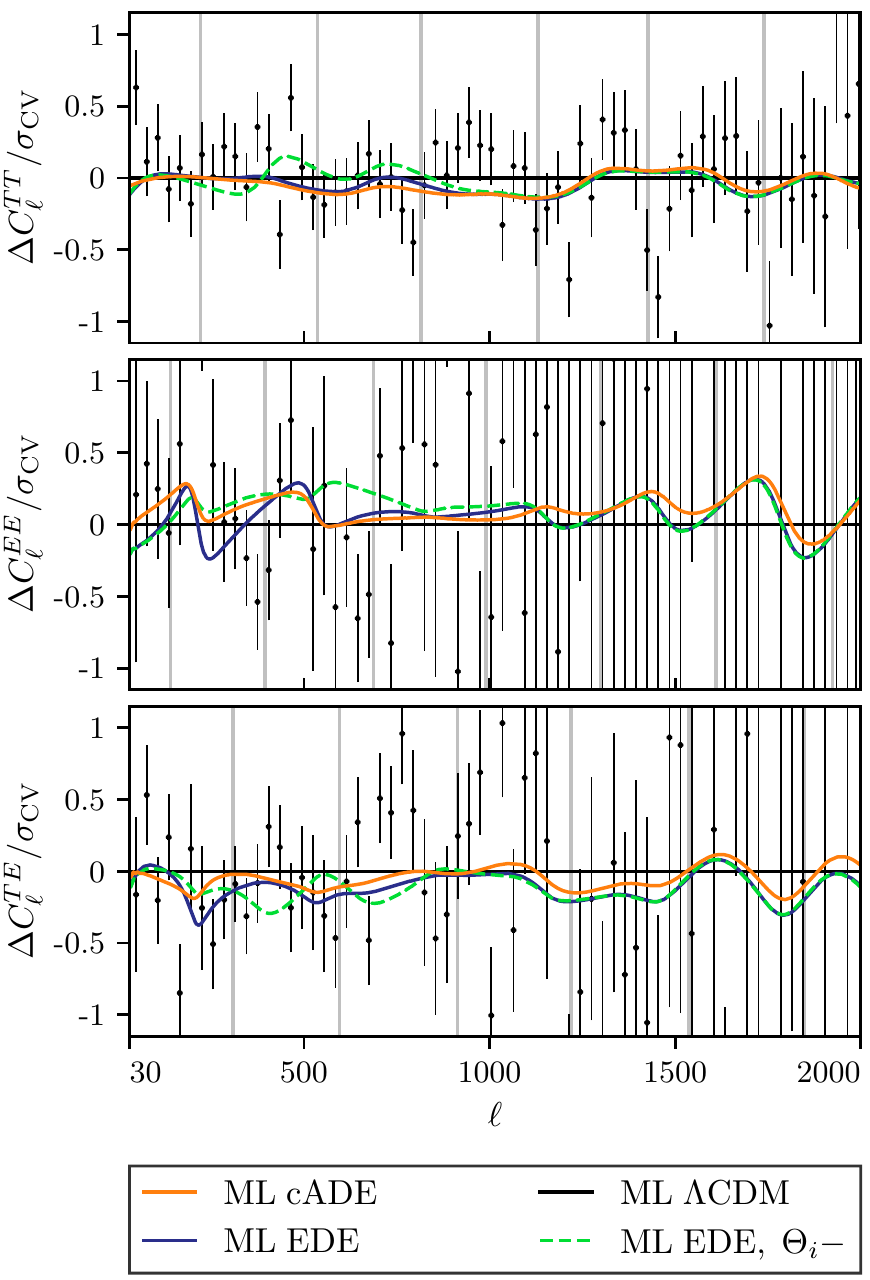}
\caption{
\label{fig:residual-cADE}
Canonical scalar field model and data residuals of ML  cADE (orange solid) and ML EDE (dark blue solid) models with respect to the ML $\Lambda$CDM model as in Fig.~\ref{fig:residual}.
The model with $\Delta\Theta_i = -0.5$ from ML EDE (green dashed) is also shown. 
}
\end{figure}

These simple canonical or cADE models still provide good fits to the data 
as illustrated in Fig.~\ref{fig:residual-cADE} for the ML cADE model of Table~\ref{tab:parameters}.
The main difference compared with ML ADE is the slight lowering of $H_0$ from 70.81 to 
70.57.  
The total improvement over $\Lambda$CDM for 2 extra parameters is $\Delta \chi^2_{\rm tot}=-12.7$ with $-3.6$ actually coming
from the improved fit to the CMB as shown in Table~\ref{tab:chi2}.
More concretely, ML cADE makes CMB lensing a little bit worse by $\Delta\chi^2_{\rm lens}=+1.1$ but fits the TT and polarization spectrum better by $\Delta(\chi^2_{\rm plik}+\chi^2_{\rm lowTEB})=-4.7$. If compared to ML $\Lambda$CDM fit to CMB only, the ML cADE fits CMB lensing as well and fits the TT and polarization spectrum better by $\Delta(\chi^2_{\rm plik}+\chi^2_{\rm lowTEB})=-1.6$.
\footnote{ We have also explicitly checked that a direct solution for the scalar field Klein Gordon equation is nearly indistinguishable from a cADE model with the best matching parameters, e.g.~$\Delta\chi^2=1.4$ for a cosmic variance limited TT, TE, EE measurement to $\ell \le 2000$.  
This holds for these gravitationally sourced, or adiabatic, field perturbations whereas modeling isocurvature fluctuations from initial field perturbations from inflation would require matching the radiation dominated evolution in the equation of state $(1+w) \propto a^4$ implied by Hubble friction through Eq.~(\ref{eqn:Hubbledrag}) \cite{Gordon:2004ez}, which Eq.~(\ref{eqn:eos}) does not do.}

\begin{figure*}
\centering
\includegraphics[width=\textwidth]{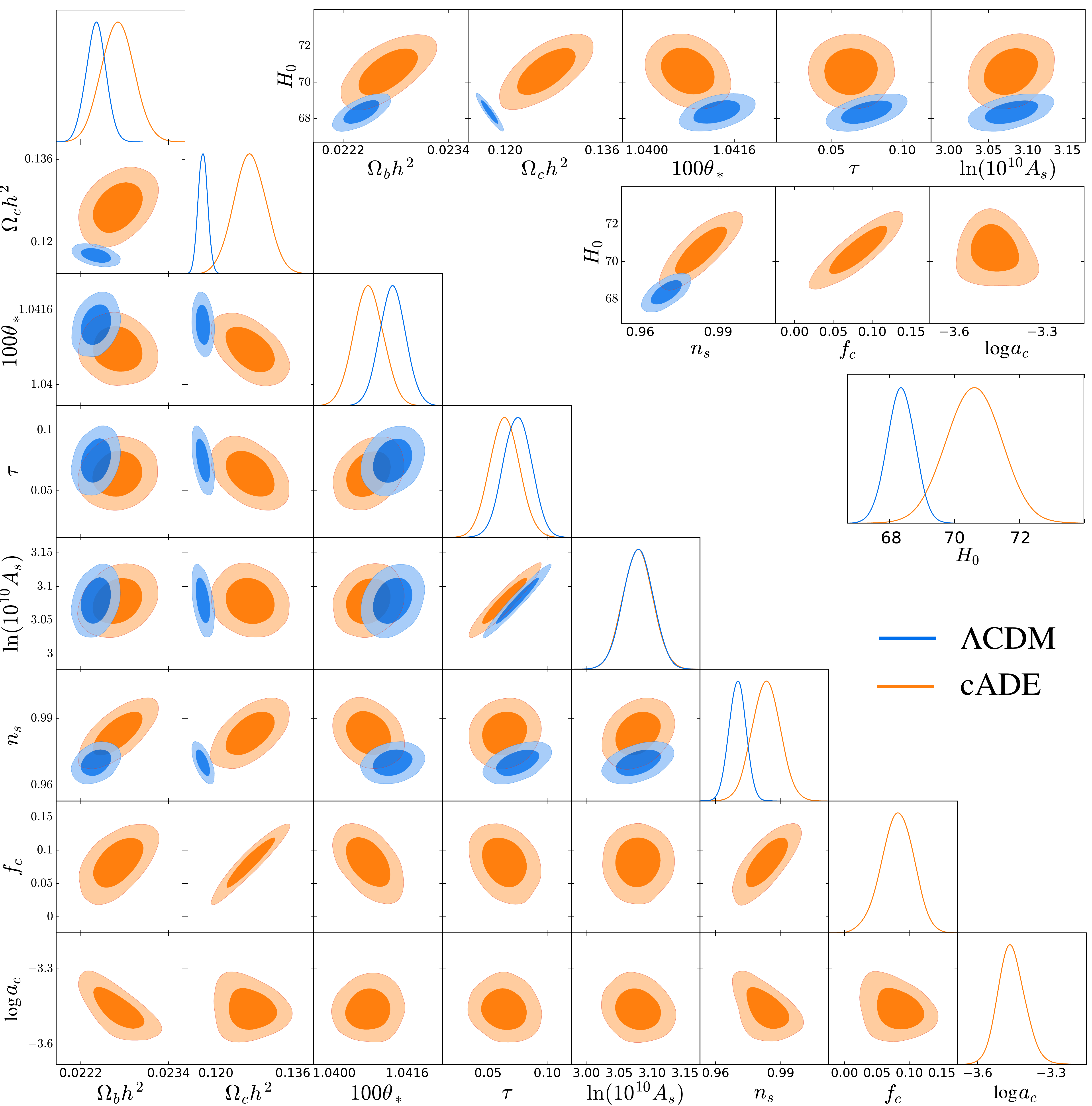}
\caption{ 
\label{fig:posterior-wn1}
The marginalized joint posterior of the parameters of the cADE model, obtained using our combined datasets. 
$\Lambda$CDM results are also added for comparison.   
The 8 fundamental parameters are shown in the lower triangle 
whereas the implications for $H_0$ are shown in the upper triangle.
The darker and lighter shades correspond respectively to the 68\% C.L. and the 95\% C.L. 
}
\end{figure*}	

\figref{fig:posterior-wn1} shows the parameter covariances and posteriors in the cADE model.  The centered values for the ML parameters is indicative of the nearly Gaussian posteriors and reflects the fact that the parameters are constrained mainly by the data rather than the priors.   The one exception is $a_c$ 
since if $a_c \rightarrow 0$ any $f_c$ is equivalent to $\Lambda$CDM so that the prior volume begins to matter.    
Even in this case, $\Lambda$CDM is sufficiently disfavored so that constraints on $a_c$ are data not prior driven.   
Correspondingly the ADE fraction is significantly detected with $f_c= 0.082\pm 0.025$.

This model also illustrates the main compensation between raising the ADE fraction  $f_c$ and raising the CDM density $\Omega_c h^2$ as well as adjusting $\Omega_b h^2$ and $n_s$ slightly higher to minimize the data residuals (see Fig.~\ref{fig:posterior-wn1}).
The change in $\theta_*$ to lower values is also notable.  
The modifications in driving make a small change in the phasing of the CMB acoustic peaks relative to its sound horizon.  
Note that in $\Lambda$CDM, $\theta_*$ drifts lower once the high multipoles $\ell > 800$ are included~\cite{Aghanim:2016sns}.

Finally under the -POL data set, the ADE canonical model allows a higher $H_0= 71.55 \pm 1.05$ and ML value of 71.93.   
This is because of the limitations the TE spectrum around $\ell \lesssim 500$ places on these solutions as discussed in \S \ref{Sec:pol}.

\section{Relation to Prior Work} \label{Sec:previous}
\begin{figure}[!ht]
\centering
\includegraphics[width=\columnwidth]{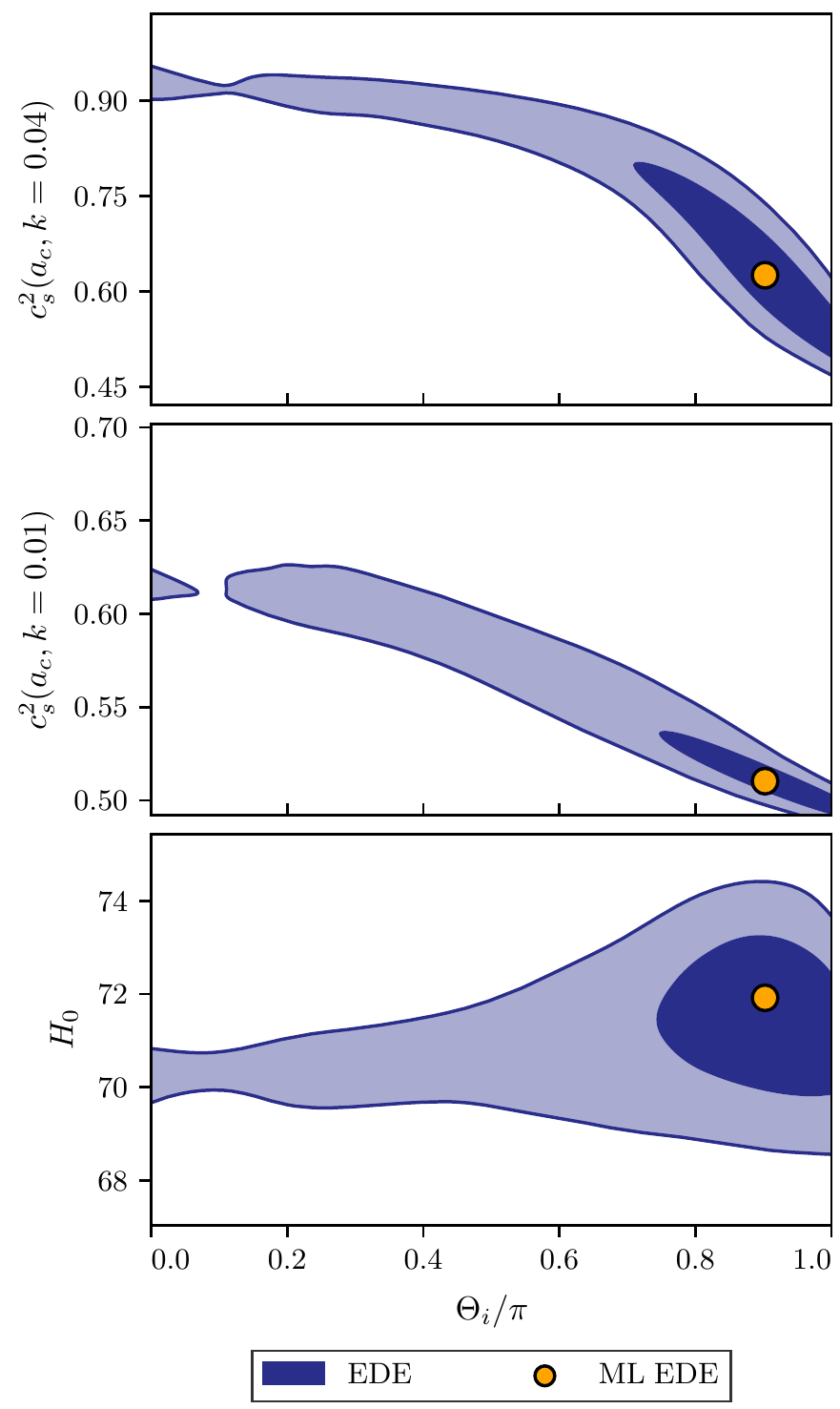}
\caption{
The marginalized distribution of the EDE initial phase $\Theta_i$ and some other parameters, obtained using our combined data set.
The darker and lighter shades correspond respectively to the 68\% C.L. and the 95\% C.L. The orange circle
indicates the maximum likelihood values for EDE. 
\label{fig:EDEphase}
}
\end{figure}		

In Ref.~\citep{Poulin:2018cxd}, a canonical scalar field component, referred to as early dark energy,  with a potential
\begin{equation} \label{eqn:poulin}
V(\phi) \propto [1-\cos(\phi/f)]^n,
\end{equation} 
 plays a similar role as our ADE. 
Unlike ADE,  EDE oscillates after being released from Hubble drag and  Ref.~\cite{Poulin:2018dzj} finds that the time averaged background equation of state can be modeled by Eq.~(\ref{eqn:eos}) with $p=1$ and
\begin{equation}
\wn = \frac{n-1}{n+1} \,.
\end{equation}
As we have discussed in the previous sections, $\wn$ is a relevant parameter for the resolution of Hubble tension, so its
adjustment should be considered a parameter variation in the EDE model in spite of $n$ taking discrete integer values.

The time-averaged behavior of perturbations is described by a fluid approximation with a  rest frame sound speed~\cite{Poulin:2018dzj}:
\begin{equation}
	c^2_{s} (a,k)= 
	\begin{cases}
	1 , & a \le a_c,  \\
	\dfrac{2a^2(n-1)\varpi^2+k^2}{2a^2(n+1)\varpi^2+k^2}  \,,
	& a > a_c \,.
	\end{cases}
	\label{eqn:EDEcs}
\end{equation}
Unlike in our case, this fluid description is approximate, especially at $a_c$.  
The time dependence of $\varpi$ is fixed by the parameters $a_c,\wn$ and the initial field position $\Theta_i=\phi_i/f$~\cite{Poulin:2018dzj}:
\begin{equation}
\varpi(a) = \frac{\mathcal{G}}{\mathcal{P}} \sqrt{\frac{6\mathcal{P}+2}{8n}\frac{\Theta_i}{\sin \Theta_i}} H(a_c) {a}^{-3\wn} \,.
\end{equation}
Here ${\mathcal P}= H(a_c) t$ and we approximate it as: 
\begin{equation}
\mathcal{P}(x=a_c/a_{\rm eq}) = \frac{2}{3} \frac{  x^2 -x + 2 \sqrt{1+x} -2}{x^2} \,,
\end{equation}
while
\begin{eqnarray}
\mathcal{G}(a_c,n) & = & \frac{\sqrt{\pi}\Gamma(\frac{n+1}{2n})}{\Gamma(1+\frac{1}{2n})} 2^{-\frac{n^2+1}{2n}} 3^{\frac{1}{2}(\frac{1}{n}-1)} a_c^{3-\frac{6}{n+1}}\nonumber\\
	& & \times  \left[a_c^{\frac{6n}{n+1}}+1\right]^{\frac{1}{2}(\frac{1}{n}-1)} \,.
\end{eqnarray}
Note that for $\wn>0$, $\varpi$ decreases with $a$, and so the sound speed evolves from $1$ at $a\le a_c$ 
back to $1$ at late times, with higher $k$ exhibiting smaller amplitude deviations, and a $k$-dependent minimum $c_s^2(a_c,k)$.

The EDE model therefore has four parameters $a_c, f_c, \Theta_i$ and $n$.  
Following Ref.~\cite{Poulin:2018dzj}, we choose the best value $n=3$, which corresponds to $\wn=1/2$, and conduct an MCMC likelihood analysis on the remaining parameters.  
We treat $a_c$ and $f_c$ as in the ADE model and impose a flat prior on $0\leq \Theta_i/\pi  \leq 1$.  Because of the large parameter volume  of degenerate models
 around $\Lambda$CDM, we only sample the posterior until $R-1 < 0.05$ which should give an adequate, but not perfect, estimate of parameter constraints out to $95\%$ C.L.
The results are compared with our ADE model in Table~\ref{tab:chi2}.  
The EDE ML model allows a slightly higher $H_0=71.92$ and hence a better fit to the data $\Delta \chi^2_{\rm tot}=-2.5$ for one extra parameter over ML ADE with $c_s^2=\wn$ or $\Delta\chi^2_{\rm tot} = - 3.9$ for two extra parameters compared with the cADE ML.   
Note that with the -POL dataset, the EDE and ADE models have comparable performances.

The main phenomenological difference between the ADE and EDE models is the parametrization of the sound speed.
The sound speed was indeed also varied in Ref.~\cite{Poulin:2018dzj} but its impact was not discussed.
As we have seen in our ADE model a low $\wn$ generally requires a low $c_s^2$. 
In the EDE model, this translates into specific requirements for the initial phase $\Theta_i$.  
In Fig.~\ref{fig:EDEphase}, we show the relationship between constraints on $\Theta_i$, $H_0$ and the minimum sound speed at $k=0.01, 0.04$.  
As we can see, achieving a higher value of $H_0$ requires a large initial phase and its ML value is $\Theta_i/\pi=0.90$.     The  68\% confidence region  is $0.72 \le \Theta_i/\pi \le 0.94$
and $\Theta_i/\pi < 1/2$ is excluded at 93\% C.L.  Note that the upper range exceeds the
value required for the validity of the fluid mapping approximation, $\Theta/\pi \sim 0.96$~\cite{Poulin:2018dzj}.

The reason for this preference is that $\Theta_i$ controls the minimum sound speed.  Following the degeneracy line in Fig.~\ref{fig:wn-cs2} to $w_n=1/2$, we would expect that for a constant sound speed $c_s^2 \approx 0.77$.  Given the effective EDE sound speed of Eq.~(\ref{eqn:EDEcs}), this represents an average over the relevant timescales and wavemodes.  
For the ML EDE model $c_s^2(a_c,0.01)=0.51$ and $c_s^2(a_c,0.04)=0.63$ as the minimum value for each $k$-mode.  

The scale dependence of the sound speed also explains the slightly better fit to CMB  data, specifically the TE data.   
In Fig.~\ref{fig:residual-cADE}, we compare the EDE and cADE residuals for their respective ML models.    Notice that the TT residuals are very similar.
However, in TE, by allowing the sound speed to decrease in the $k$ range associated with $\ell < 500$, the EDE model exposes more of the driving reduction at $\ell \sim 200$ from raising $\Omega_c h^2$ as discussed above but now without adverse consequences elsewhere.
This in turn better fits the low TE residuals and allows $H_0$ to increase further relative to the ADE model.     
We also show the impact of reducing $\Theta_i$ in the ML EDE model, making the sound speed closer to one at all times.  The most significant effect is localized to  $\ell < 500$ 
and in particular destroys the pattern of lower TE at $\ell \sim 200$ vs $\ell \sim 400$ compared with cADE.

We conclude that the small improvement of the EDE over ADE fit requires a specific range in the initial phase that lowers the sound speed in a scale-dependent way. 
Comparing to the canonical ADE mode, this improvement gives $\Delta\chi^2_{\rm tot} = - 3.9$ for two extra parameters $w_n, \Theta_i$ and is therefore marginal.     
In the future, polarization measurements that approach the cosmic variance limit can
distinguish between the EDE and ADE classes.   For example, we forecast that
with cosmic variance TT,TE,EE measurements to $\ell\le 2000$, the current best fit EDE model
differs from the closest ADE model by $\Delta\chi^2 = 22.4$.
Furthermore the ADE model provides a general class of exact solutions where the potential energy is converted quickly to kinetic, whereas the EDE model requires a specific set of initial conditions to achieve a similar phenomenology with an approximation to an oscillating field.

Relatedly, Ref.~\cite{Agrawal:2019lmo} considers a model where the scalar field oscillates in a monomial potential
\begin{equation}
V(\phi) \propto \phi^{2 n} \,,
\end{equation}
with parameters adjusted to reproduce the EDE phenomenology.
This coincides with Eq.~(\ref{eqn:poulin}) only near the bottom of the potential, $\Theta_i \ll \pi/2$ where the potential is convex rather than concave.
As pointed out in Ref.~\cite{Agrawal:2019lmo} the model has significantly worse performances than the EDE model.  We identify here that this is related to the  initial field being in the concave rather than the convex part of the potential which raises the sound speed.

Finally, while this work was nearing completion, Ref.~\cite{Alexander:2019rsc} proposed that a fast-roll or kinetic energy dominated period in a two-field model might relieve the Hubble tension.

\begin{table}[!ht]
\centering
\begin{tabular}{@{}lcllcr@{}}
\toprule
\multicolumn{1}{@{}l}{Model\,(Data)} &
\multicolumn{1}{c}{$\Delta N $} &
\multicolumn{1}{c}{ $H_0$} &
                   $\Delta\chi^2_{\rm tot}$  & $\Delta\chi^2_{\rm CMB}$  &
\multicolumn{1}{c}{$\Delta\chi^2_{\rm H0}$}                    
                   \\
\toprule
cADE        & 2              & 70.57(70.60$\pm$0.85)  \, & -12.7              \,      & -3.6                      & -8.8  \\
ADE            & 3$^*$,4              & 70.81(70.20$\pm$0.88)  & -14.1                     & -3.7                      & -9.6  \\
EDE            & 4              & 71.92(71.40$\pm$1.09)  & -16.6                     & -3.7                      & -12.5  \\
\colrule
cADE(-POL)   & 2              & 71.93(71.55$\pm$1.05)  & -12.8                     & -0.4                      & -11.2  \\
ADE(-POL)       & 3$^*$,4              & 72.27(71.30$\pm$1.03)  & -15.1                     & -2.4                      & -11.8  \\
EDE(-POL)      & 4              & 72.40(72.35$\pm$1.25)  & -15.9                     & -2.9                      & -12.1  \\
\botrule
\end{tabular}
\caption{
\label{tab:chi2}
$H_0$ results for the ML cADE, ADE and EDE  models and posterior constraints with the joint data set and with CMB polarization data removed (-POL).   
$\Delta N$ is the number of additional parameters in addition to the $\Lambda$CDM ones.  
$^*$Note that ML ADE in the potential conversion case where $c_s^2=\wn$, is essentially the same as the general case  but with $\Delta N=3.$
The total $\Delta\chi^2_{\rm tot}$ relative to the $\Lambda$CDM model is broken down into contributions from the Planck CMB data sets and the local $H_0$ measurement.  
}
\end{table}

\section{Discussion} \label{Sec:discussion}
Acoustic dark energy, appearing around the epoch of matter radiation equality, can substantially relieve the tension between CMB inference of $H_0$ and local measurements, exhibited in the $\Lambda$CDM model.   
The presence of extra energy density lowers the CMB sound horizon that anchors the inverse distance ladder for BAO and SN, while its disappearance before and after equality allows for a good fit to CMB data in the damping tail.  
Furthermore by introducing ADE at equality, the gravitational effects of raising the cold dark matter density can be balanced by the acoustic oscillations in the ADE itself. 

\begin{figure}[!ht]
\centering
\includegraphics[width=0.49\textwidth]{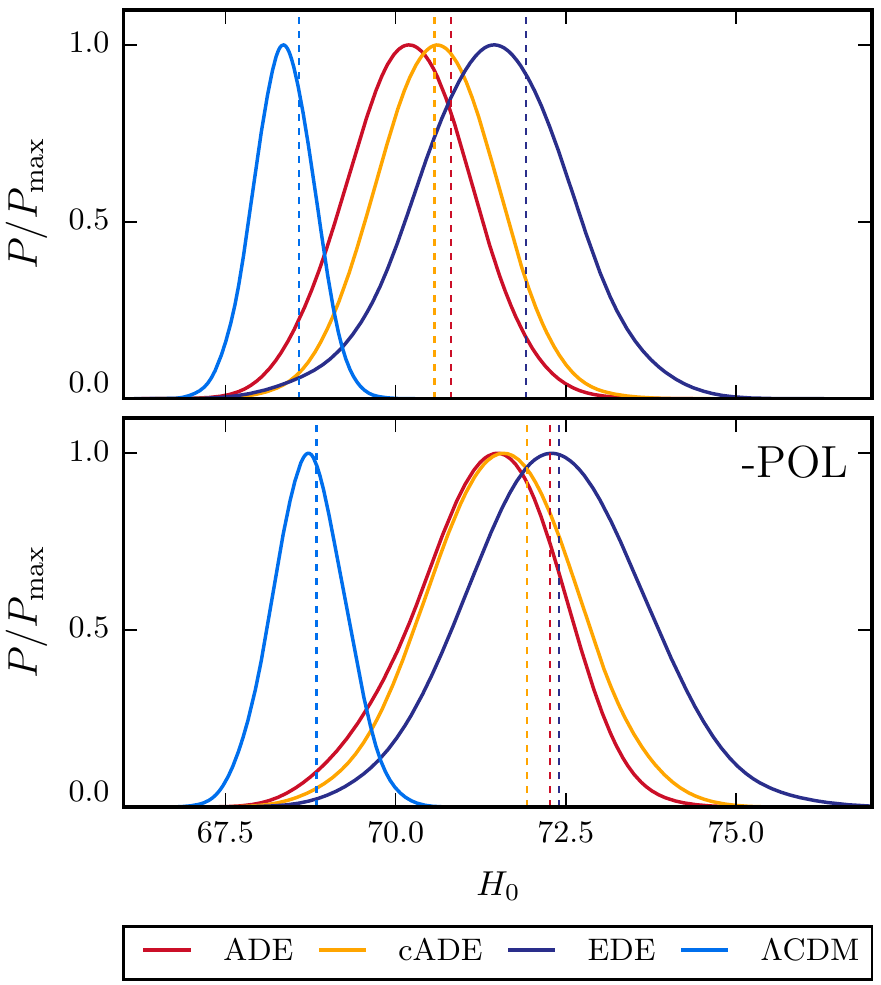}
\caption{
The marginalized posterior distribution of the $H_0$ parameter in the four models considered in Table~\ref{tab:chi2} for two different datasets: our combined data set  (upper panel) and the same with CMB polarization data removed, -POL (lower panel). 
The dashed vertical lines indicate the ML values for different models.
\label{fig:H0}
}
\end{figure}

Our main findings regarding the Hubble tension are summarized in Table~\ref{tab:chi2} and Fig.~\ref{fig:H0}.  
In all cases relieving the Hubble tension requires ADE to be a $\sim 8\%$ contribution to the total energy density around matter radiation equality, leading to at least a two parameter extension to $\Lambda$CDM.  

In the general ADE class of models, the acoustic phenomenology is controlled by two additional parameters, the asymptotic equation of state $\wn$ at late times, and the sound speed $c_s^2$.
The sound speed plays a crucial role in the gravitational driving of CMB acoustic oscillations through its impact on the Weyl potential, leading to a strong correlation between the two, consistent with $\wn = c_s^2 \approx 1$ around the maximum likelihood model.    

Fixing these two parameters to the canonical model, cADE, leads to only a minor degradation in the ability of ADE to relax the Hubble tension in the ML model, $H_0=70.57$ vs.~$H_0 = 70.81$.
In fact, the two parameter model has the advantage of providing a more significant detection of the ADE fraction $f_c=0.082\pm 0.025$ and thus allows less parameter volume around the $\Lambda$CDM limit, producing a posterior centered around the ML:  $H_0 = 70.60 \pm 0.85$ in cADE vs.~the lower $H_0 = 70.20\pm 0.88$ shift in ADE.
  
Note that the fit to the CMB data themselves improves and combined with the higher $H_0$ provided for an improvement of $\Delta \chi^2_{\rm tot} =-14.1$ for
3 parameters ($\wn=c_s^2$) and $\Delta \chi^2_{\rm tot} = -12.8$ for 2 parameters ($\wn=c_s^2=1$) respectively.

This class of $\wn = c_s^2 \sim 1$ ADE models corresponds to scalar fields which convert their potential to kinetic energy efficiently around their release from Hubble drag.  By setting this epoch to be around matter radiation equality, we obtain a robust mechanism for relieving the Hubble tension in a wide class of potentials.

For canonical scalar fields, we explicitly determine the requirements on the potential: that its slope allows for Hubble drag release around equality where its amplitude is set to the $\sim 8\%$ fraction required by the data.   
Any potential that obeys this property and efficiently converts potential to kinetic energy until the latter redshifts away will satisfy these
requirements.   
As a proof of principle, we explicitly construct an example where the potential is locally quadratic around its release.    
In this model, the timing of the release to equality is not explained but the identification of this coincidence may lead to more sophisticated models where it is.  

The robustness and generality of this potential-kinetic conversion mechanism for relieving the Hubble tension separates it from similar models in the literature.   
In Refs.~\cite{Poulin:2018cxd,Agrawal:2019lmo}, the EDE scalar field oscillates after Hubble drag release leading to an effective fluid described by time-averaged values of $\wn$ and $c_s^2$.  
Converting our requirements on the  relationship between the two, we find that in Ref.~\cite{Poulin:2018cxd}, where the potential is periodic, the initial field must be on the concave part of the potential, and near the maximum to best relieve the Hubble tension.   
This also explains the poorer fits in Ref.~\cite{Agrawal:2019lmo}, where the potential is convex and matches the periodic potential only near the minimum.   

The periodic EDE model~\cite{Poulin:2018cxd,Agrawal:2019lmo} allows for a slightly higher $H_0 = 71.40 \pm 1.09$ and better $\Delta \chi^2_{\rm tot}=-16.6$ for four parameters as compared with ADE. 
Most of this improvement comes from the fit to the Planck polarization data.   
As shown in Fig.~\ref{fig:H0}, without these data, EDE and ADE perform similarly at ML with $H_0=72.27$ vs $72.40$  respectively. 
The reason is that between equality and recombination, changes in acoustic driving between raising the CDM density and adding the dark component no longer cancel.   
This is in fact a beneficial feature of both models since the Planck TE data show low residuals with respect to the ML $\Lambda$CDM model around $\ell \sim 200$. 
By allowing the effective sound speed to depend on wavenumber, the EDE model fits this region better without violating constraints elsewhere.  However, when compared with our  cADE model this extra improvement of $\Delta \chi^2_{\rm tot}\approx 3.9$   comes at the cost of two extra parameters and a less robust mechanism for relieving Hubble tension.

Finally, in all cases, the predicted deviations in the EE power spectrum, while not at a level testable by Planck data, are highly
significant compared with cosmic variance.   
Future polarization data can provide key tests for these and other dark component explanations of the Hubble tension.  

\acknowledgments

We thank 
Nicola Bartolo,
Daniel Grin,
Tanvi Karwal,
Macarena Lagos,
Michele Liguori,
Samuel Passaglia,
Vivian Poulin
for useful discussions.
MXL, WH and MR  are supported by U.S.~Dept.~of Energy contract DE-FG02-13ER41958 and the Simons Foundation.  Computing resources were provided by the University of Chicago Research Computing Center through the Kavli Institute for Cosmological Physics at the University of Chicago. GB acknowledges financial support from Fondazione Ing. Aldo Gini. 

\bibliography{biblio}

\end{document}